\documentclass[revtex4,11pt]{article}
\usepackage[UKenglish]{babel}
\usepackage {CJK}
\usepackage{amssymb}
\usepackage{mathrsfs}
\usepackage{dsfont}
\usepackage{amsmath,amsthm}
\usepackage{amsfonts}
\usepackage{enumerate, xspace}
\usepackage{xcolor}
\RequirePackage{hyperref}
\usepackage[bottom,first]{draftcopy}
\usepackage{changebar}
\usepackage{hyperref}
\usepackage{pdfsync}
\usepackage{graphicx}
\usepackage{mathtools}
\usepackage{verbatim}
\usepackage{a4wide}
\usepackage{pgfplots}
\usepackage{relsize}
\usepackage{bbm}
\usepackage{systeme}
\usepackage[labelfont=bf]{caption}
\usepackage{mathtools}
\usepackage[md]{titlesec}
\usepackage{tikz}
\usepackage{pgfplots}
\usepackage[toc,page]{appendix}
\usepackage[left=2.5cm,right=2.5cm]{geometry}

\usepgfplotslibrary{fillbetween}
\usetikzlibrary{patterns}

\numberwithin{equation}{section}

\usepackage{amsthm}

\newtheorem{thm}{Theorem}[section]
\newtheorem{lem}[thm]{Lemma}

\newtheorem{prop}[thm]{Proposition}

\newtheorem*{lem*}{Lemma}

\newcommand\ve{\varepsilon}
\newcommand\vf{\varphi}

\newcommand\nn{\nonumber}

\pgfplotsset{compat=1.12}

\titleformat*{\section}{\bf\Large}
\titleformat*{\subsection}{\bf\large}
\titleformat*{\subsubsection}{\bf}

\begin{document}

\begin{center}
{\bf\LARGE Fick's Law in Non-Local Evolution Equations} \\
\end{center}
\begin{center}
{\Large Roberto Boccagna} \\ 
\vspace{2mm}
{Gran Sasso Science Institute, Viale F. Crispi 7, 67100 L'Aquila, Italy\\
{\hspace{-5.49cm}e-mail: \texttt{roberto.boccagna@gssi.it}}}
\end{center}

\vspace{3mm}

\begin{abstract}
We study the stationary non-local equation which corresponds to the energy functional of a one-dimensional Ising spin system, in which particles interact via a Kac potential. In particular, following some of the techniques developed in \cite{DMPT}, we construct a time invariant profile for the model proposed in \cite{GIAC}. The boundary conditions share the same sign and both lie above the value $m^*\left(\beta\right)=\sqrt{1-1/\beta}$, which divides the metastable region from the unstable one, the inverse temperature being fixed and larger than the critical value $\beta_c=1$. Due to the non-equilibrium setting, a non zero magnetization current, which scales with the inverse of the size of the volume $\ve^{-1}$, do flow in the system \cite{CDMP}. Here $\ve^{-1}$ also represents the ratio of macroscopic and mesoscopic length. We show that for $\ve>0$ small enough, the stationary profile has no discontinuities so that no phase transition occurs; although expected when the magnetizations are larger than $m_{\beta}$, this turns out to be non trivial at all in the metastable region. Moreover, when $\ve^{-1}\to\infty$, the solution converges to that of the corresponding macroscopic problem, i.e. the local diffusion equation. The validity of the Fick's law in this context is then established. 
\end{abstract}

\section{Introduction}

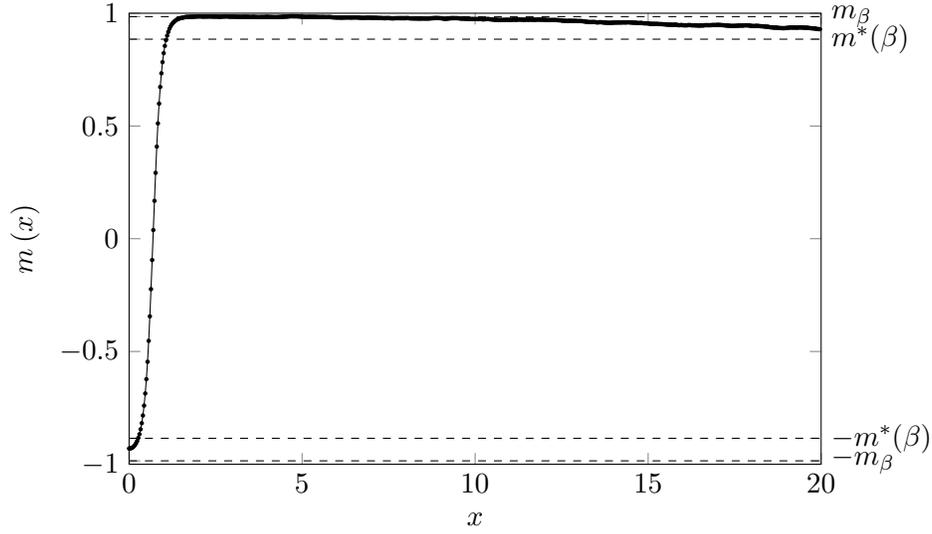
\begin{figure}[htbp!]
\centering
\begin{tikzpicture}[trim axis left, trim axis right]
\begin{axis}[
width=9.2cm,
height=6cm,
scale only axis,
ylabel=$m\left(x\right)$,
xlabel=$x$,
ymin=-1.0,
ymax=1.0,
xmin=0,
xmax=20,
xtick={0,5,10,15,20},
ytick={-1,-0.5,0,0.5,1},
extra y ticks={-0.985,-0.885,0.885,0.985},
extra y tick labels={$-m_{\beta}$,$-m^*(\beta)$,$m^*(\beta)$,$m_{\beta}$},  
extra y tick style={y tick label style={right, xshift=9.2 cm}},
]
\addplot[thin,mark=*,mark size=0.225mm,mark options={color=black,line width=0.03pt,fill=black},color=black]
table{data93_8SX.dat};
\draw[dashed,color=black] (axis cs:0,0.885) -- (axis cs:20,0.885);
\draw[dashed,color=black] (axis cs:0,0.985) -- (axis cs:20,0.985);
\draw[dashed,color=black] (axis cs:0,-0.885) -- (axis cs:20,-0.885);
\draw[dashed,color=black] (axis cs:0,-0.985) -- (axis cs:20,-0.985);
\end{axis}
\end{tikzpicture}
\caption{Steady magnetization profile resulting from running the stochastic microscopic dynamics as defined in \cite{CDMP}. As in the mentioned work, here $\beta=2.5$, $\gamma^{-1}=30$, $L=600$. The infinite reservoirs are at constant values $\mu^-=-0.93$ (left), $\mu^+=0.93$ (right). In one mesoscopic unit the magnetization jumps from $\mu_-$ to reach its maximum.}\label{fig}
\end{figure}

The problem of characterizing stationary steady currents is widely addressed in the context of Non-Equilibrium Statistical Mechanics \cite{JL1,JL2,JL3,DLSS}. Many results have been obtained concerning steady states with a non zero current when no phase transition occurs. Nevertheless very few is known, at least mathematically, on  the most general case in which, in the thermodynamic limit, sharp discontinuities may connect regions of different phases. 
\newline
Non trivial macroscopic effects induced by non-local interactions are known from the 40's, when an anti-gradient diffusion was observed in a pioneering experiment performed on weld metals \cite{LSD}. These evidences have been recently mimicked by running numerical simulations of a discrete-time 1$d$ stochastic cellular automaton, in which particles interact through a long range potential staying in contact with two infinite reservoirs. When the prescribed boundary densities lie in the metastable region and are in different phases, the system rearranges in such a way that the current flows from the reservoir at lower density to that at the higher one (\textit{uphill diffusion}); this seems to be connected to the formation of \textit{bumps}, which turn in boundary layers in the infinite volume limit \cite{CDMP}. The typical stationary profile is similar to that in Figure \ref{fig}: depending on the starting configuration, one of the boundaries jumps to the opposite metastable phase in one mesoscopic unit; in any case, the current is positive, following the slope of the quasi-linear part. 


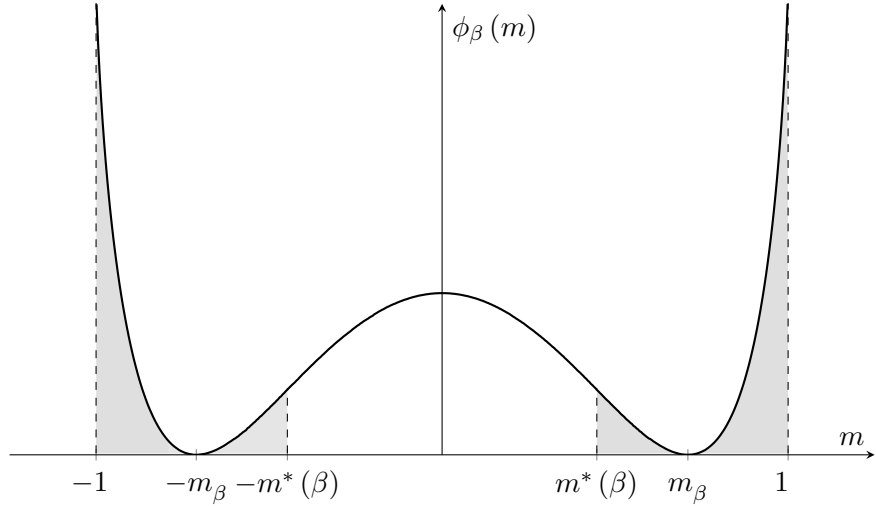
\begin{figure}[htbp!]
\centering
\begin{tikzpicture}[trim axis left, trim axis right]
\def\doublewell{\x,{-(\x^2/2)+0.8*((1-\x)/2)*ln((1-\x)/2)+0.8*((1+\x)/2)*ln((1+\x)/2)+1.2216}}
\def\zerocurve{\x,{0}}
\begin{axis}[
width=11.5cm,
height=6cm,
scale only axis,
axis lines=middle,
inner axis line style={=>},
ylabel=$\phi_{\beta}\left(m\right)$,
xlabel=$m$,
xmin=-1.25,
xmax=1.25,
domain=-1.1:1.1,
xtick={5},
ytick={-1},
extra x ticks={-1,-0.710412,-0.4472,0.4472,0.710412,1},
extra x tick labels={$-1^{\textcolor{white}{*}}$,$-m_{\beta}^{\textcolor{white}{*}}$,$-m^*\left(\beta\right)$,$m^*\left(\beta\right)$,$m_{\beta}^{\textcolor{white}{*}}$,$1^{\textcolor{white}{*}}$},  
]
\addplot[thick,color=black,samples=500,name path=p]
{-(x^2/2)+0.8*((1-x)/2)*ln((1-x)/2)+0.8*((1+x)/2)*ln((1+x)/2)+1.2214};
\path[name path=axisp] (axis cs:0.4472,0.6384) -- (axis cs:1,0.6384);
\addplot [
thick,
pattern=north east lines,
color=gray,
fill opacity=0.25
]
fill between[
of=p and axisp,
soft clip={domain=0.4472:1},
 ];
\path[name path=axism] (axis cs:-1,0.6384) -- (axis cs:-0.4475,0.6384);
\addplot [
thick,
pattern=north east lines,
color=gray,
fill opacity=0.25
]
fill between[
of=p and axism,
soft clip={domain=-1:-0.79},
 ];
 \addplot [
thick,
pattern=north east lines,
color=gray,
fill opacity=0.2
]
fill between[
of=p and axism,
soft clip={domain=-0.79:-0.4472},
 ];
\draw[dashed,color=black] (axis cs:0.4472,0.63) -- (axis cs:0.4472,0.65);
\draw[dashed,color=black] (axis cs:-0.4472,0.63) -- (axis cs:-0.4472,0.65);
\draw[dashed,color=black] (axis cs:1,0.63) -- (axis cs:1,0.8);
\draw[dashed,color=black] (axis cs:-1,0.63) -- (axis cs:-1,0.8);
\end{axis}
\end{tikzpicture}
\caption{The double well potential $\phi_{\beta}\left(m\right) = -{m^2\over 2}-{1\over\beta}S_{\beta}\left(m\right)$. The region between $-m^*\left(\beta\right)$ and $m^*\left(\beta\right)$ (in white) is unstable and will be avoided.} 
\end{figure}\label{dw}


\noindent
In the macroscopic limit, the jumping interval shrinks into a point, so that a singularity arises: thus, when the limits are performed from the interior of the bulk, it is like the system undergoes the influence of magnetizations of the same sign.
\newline
Kac potentials provide a natural instrument for modeling long range, non-local interactions in magnetic spin systems. Unlike classical mean field approximations, indeed, particles are affected in a weakly, decreasing way by spins located at large distances, the typical interaction length scaling with the inverse of a parameter $\gamma$ which goes to zero in the mesoscopic limit \cite{EP}. In this context, the purpose would be to built a general theory and prove the existence of bumps when opposite metastable boundary conditions are imposed. However, the starting point for such an analysis should be represented by the case in which the boundary magnetizations are constant, both positive (negative) and larger than $m^*\left(\beta\right)$. As mentioned above, this is strongly motivated by the fact that as the size of the system increases, the two solutions get closer one to the each other. In these hypothesis, even though we expect the Fick's law to be valid in the stable regime, i.e. for values larger than $m_{\beta}$ as provided by the axiomatic theory, we prove here that it actually holds true even when the boundaries are strictly metastable.
\newline
We thus study the integro-differential equation corresponding to the one dimensional Lebowitz-Penrose free energy functional of the model introduced in \cite{GIAC}:
\begin{eqnarray}
&&\mathcal{F}_{\beta,\Lambda_{\ve}}\left[m\mid m_{\Lambda_{\ve}^c}\right] =
\mathcal{F}_{\beta,\Lambda_{\ve}}\left[m\right] + {1\over 2} \int_{\Lambda_{\ve}} \int_{\Lambda_{\ve}^{c}} J\left(x,y\right) \left[m\left(x\right)-m_{\Lambda_{\ve}^c}\left(y\right)\right]^2\mathrm{d}x\,\mathrm{d}y  \\
&&\mathcal{F}_{\beta,\Lambda_{\ve}}\left[m\right] = 
-{1\over 2} \int_{\Lambda_{\ve}} m^2\left(x\right)\mathrm{d}x -{1\over\beta}\int_{\Lambda_{\ve}}S_{\beta}\left(m\left(x\right)\right) \mathrm{d}x 
- {1\over 2}\int_{\Lambda_{\ve}}\int_{\Lambda_{\ve}} J\left(x,y\right) \left[m\left(x\right)-m\left(y\right)\right]^2\mathrm{d}x\,\mathrm{d}y \nn \\ \\
&& S_{\beta}\left(m\right) = -{1+m\over 2}\log\left({1+m\over 2}\right) - {1-m\over 2}\log\left({1-m\over 2}\right)
\end{eqnarray}
where $\Lambda_{\ve}\subseteq\mathbb{R}$ is the interval $\Lambda_{\ve}\coloneqq[0,\ve^{-1}]$, $m\in L^{\infty}\left(\Lambda_{\ve}\right)$ and $m_{\Lambda_{\ve}^c}\in L^{\infty}\left(\Lambda_{\ve}^c\right)$.
\newline
Given the probability kernel:
$$ \widetilde{J}\left(x,y\right) = \widetilde{J}\left(0,y-x\right), \qquad \widetilde{J}\left(0,x\right) \coloneqq \left(1-\left|x\right|\right){{\mathbf{1}}}_{\left|x\right|\le 1}$$
$J\left(x,y\right)$ is defined, in the distributional sense, as:
\begin{equation}
J\left(x,y\right) \stackrel{\mathcal{D}'}{\coloneqq} \widetilde{J}\left(x,y\right)\mathbf{1}_{0\le x\le\ve^{-1}} + a_-\left(x\right) \delta_0\left(y\right) + a_+\left(x\right)\delta_{\ell}\left(y\right)
\end{equation}
being:
\begin{equation}
a_-\left(x\right) \coloneqq \int_{-1}^{0} \widetilde{J}\left(x,y\right) \text{d}y, \qquad
a_+\left(x\right) \coloneqq \int_{\ve^{-1}}^{\ve^{-1}+1} \widetilde{J}\left(x,y\right) \text{d}y.
\end{equation}
One basic ansatz of the axiomatic non equilibrium theory is that the steady current $j$ which flows in system is related to the free energy by \cite{DMPT,EP}:
\begin{equation}\label{flow}
j = - \ve^{-1} \chi\left(m\left(x\right)\right) {\mathrm{d}\over\mathrm{d}x} {\delta\mathcal{F}_{\beta,\Lambda_{\ve}}\over\delta m\left(x\right)}\left[m\mid m_{\Lambda_{\ve}^c}\right]
\end{equation}
so that, once denoted the magnetic susceptibility (mobility coefficient) as:
\begin{equation}
\chi\left(m\right) \coloneqq \beta \left(1-m^2\right)
\end{equation}
one gets from \eqref{flow}:
\begin{equation}\label{pb}
- {\text{d}m\over \text{d}x}\left(x\right) + \beta\left(1-m^2\left(x\right)\right) \int_{\Lambda_{\ve}} J\left(x,y\right) {\text{d}m\over \text{d}y}\left(y\right) \text{d}y = j\ve \qquad x\in\Lambda_{\ve}
\end{equation}
with boundary conditions:
\begin{equation}
m\left(0\right)=\mu^- \qquad m\,(\ve^{-1})=\mu^+.
\end{equation}
Without loss of generality, we consider here the case in which $\mu_- > \mu_+ > m^*\left(\beta\right)$, being the complementary one similar at all (see Figure \ref{dw}).
\newline
Defined the auxiliary magnetic field $h\left(x\right)$ through:
\begin{equation}
\chi\left(m\left(x\right)\right) {\mathrm{d}h\over\mathrm{d}x}\left(x\right) = -j\ve
\end{equation}
it is worth splitting the problem \eqref{pb} in two coupled equations:
\begin{equation}\label{pbs}
\begin{dcases}
m\left(x\right) \hspace{-2.5mm}&= \tanh\Big\{\beta\Big[\left(J\ast m\right)\left(x\right) + h\left(x\right)\Big]\Big\}\\
h\left(x\right) \hspace{-2.5mm}&= T\left(m\left(x\right)\right)=\widetilde{h}- j\ve \int_0^{x}\chi^{-1}\left(m\left(y\right)\right) \text{d}y
\end{dcases}
\end{equation}
with the same boundary conditions as in \eqref{pb}, while the constant of integration $\widetilde{h}$ is set to the value: 
\begin{equation}\label{htilde}
\widetilde{h} = {1\over\beta} \, \mathrm{arc}\tanh\,(\mu^-)-\mu^-.
\end{equation}
The main point is to prove that \eqref{pb} admits a $C^{\infty}\left([0,\ve^{-1}]\right)$ solution, being intended that, here and in the sequel, we say that $m\in C^{\infty}\left([0,\ve^{-1}]\right)$ if $m\in C^{\infty}\left((0,\ve^{-1})\right)$ and $m\in C^{\infty}\left(0^+\right)$, $m\in C^{\infty}\left((\ve^{-1})^{-}\right)$.
The approach we have in mind is based on a recursive strategy and is quite similar to that used in \cite{DMPT} where a diffusion problem has been examined as well. Despite these schemes share the same logic, both consisting in the change of variables \eqref{pbs} and, as we shall see, in constructing by suitable recursions a fixed point for the problem, the working assumptions and then the technical issues are different or new. In particular, in \cite{DMPT} the symmetries play a fundamental role, so that the magnetization and the other relevant quantities are antisymmetric in the considered domain. Moreover, while $j$ and $\widetilde{h}$ are actually fixed, the boundary magnetizations are in that case independent on these values and the convolutions are defined in terms of Neumann conditions. We will come back more in detail on this point further on.

\section{Main Results}

From now on, $I_{\beta} \coloneqq \left(m^*\left(\beta\right),1\right)$. We shall prove the following:
\begin{thm}\label{theorem}
There is $\bar{\ve}>0$ such that for any $\ve<\bar{\ve}$ the problem:
\begin{equation}\label{pbd}
\begin{dcases}
\;-{\mathrm{d}m\over \mathrm{d}x} \left(x\right) + \beta\left(1-m^2\left(x\right)\right) \int_0^{\ve^{-1}} J\left(x,y\right) {\mathrm{d}m\over \mathrm{d}y}\left(y\right) \mathrm{d}y = j\ve \qquad &x\in[0,\ve^{-1}] \\
\;m\left(0\right) = \mu^- \\ 
\;m\,(\ve^{-1}) = \mu^+
\end{dcases}
\end{equation}
admits a $C^{\infty}\left([0,\ve^{-1}]\right)$ solution for any couple of boundary conditions $\left(\mu^-,\mu^+\right) \in I_{\beta}\times I_{\beta}$. 
\newline
Moreover:
\begin{equation}
\lim_{\ve\,\downarrow\,0} \|m-m_0\|_{\infty} = 0
\end{equation}

\noindent
where $m_0$ solves the corresponding macroscopic problem: 
\begin{equation}\label{pb0}
\begin{dcases}
\;-{\mathrm{d}\over \mathrm{d}x} m_0\left(x\right) = {j\over 1-\beta\left(1-m_0^2\left(x\right)\right)} \ve\qquad x\in(0,\ve^{-1}) \\
\;m\left(0\right) = \mu^- \\
\;m\,(\ve^{-1}) = \mu^+.
\end{dcases}
\end{equation}
\end{thm}

\subsection{Outline of the proof}

Most of the results presented here will be discussed and systematically proved in the following sections.

\begin{center}
\textit{The iterative scheme}.
\end{center}
Once rewritten \eqref{pbd} in the form of the coupled equations \eqref{pbs}, the idea consists in defining a feasible sequence of iterated pairs $\left(m_n,h_n\right)$ which converges to the solution of \eqref{pbd}. The corresponding estimates will be performed in a proper weighted norm, whose introduction turns to be useful in order to take advantage, in the computations, of the finite range nature of the interactions.
\newline
The starting element will be the couple $\left(m_0,h_0\right)$ which satisfies the mean field type equation:
\begin{equation}\label{noj}
m_0\left(x\right) = \tanh\Big\{\beta\Big[m_0\left(x\right) + h_0\left(x\right)\Big]\Big\}
\end{equation}
where, by definition:
\begin{equation}\label{h0}
h_0\left(x\right) = \widetilde{h} - j\ve \int_0^x \chi^{-1}\left(m_0\left(y\right)\right) \text{d}y
\end{equation}
since $m_0$ solves \eqref{pb0} and $\widetilde{h}$ is given by \eqref{htilde}. 
\newline
As we shall see, this choice is motivated by the the fact that $m_0$ differs from the solution of \eqref{pbs} by a term which is of order $\ve$ in the sup norm; thus, $m_0$ almost represents a fixed point for \eqref{pb}.
\newline
The first iteration is induced by the following:
\begin{prop}\label{p2.2}
For any $\ve<\overline{\ve}$ there exists $m_1\in C^{\infty}([0,\ve^{-1}])$ which solves:
\begin{equation}\label{m1}
m_1\left(x\right) = \tanh\Big\{\beta\Big[\left(J\ast m_1\right)\left(x\right) + h_0\left(x\right)\Big]\Big\} \qquad x\in[0,\ve^{-1}],
\end{equation}
$h_0$ as in \eqref{h0}, with boundary conditions $m_1\left(0\right) = \mu^-_1$ and $m_1\,(\ve^{-1}) = \mu^+_1$, $(\mu^-_1,\mu^+_1)\in I_{\beta}\times I_{\beta}$.
\end{prop}
\noindent
Given $m_1$ solution of \eqref{m1}, one defines $h_1$ as:
\begin{equation}\label{h1}
h_1\left(x\right) = \widetilde{h} - j\ve \int_0^x \chi^{-1}\left(m_1\left(y\right)\right) \text{d}y
\end{equation}
in order to get the new pair $\left(m_1,h_1\right)$. The following result explicitly defines the whole recursion.
\begin{prop}\label{p2.3}
For any $\ve<\overline{\ve}$ and $n\in\mathbb{N}^*$ there exists $m_n\in C^{\infty}([0,\ve^{-1}])$ which solves:
\begin{equation}\label{mn}
m_n\left(x\right) = \tanh\Big\{\beta\Big[\left(J\ast m_n\right)\left(x\right) + h_{n-1}\left(x\right)\Big]\Big\}\qquad x\in[0,\ve^{-1}]
\end{equation}
with:
\begin{equation}\label{hn}
h_{n-1}\left(x\right) = \widetilde{h} - j\ve \int_0^x \chi^{-1}\left(m_{n-1}\left(y\right)\right) \mathrm{d}y,
\end{equation}
$m_n\left(0\right) = \mu^-_n$ and $m_n\,(\ve^{-1}) = \mu^+_n$, $(\mu^-_n,\mu^+_n)\in I_{\beta}\times I_{\beta}$.
\end{prop}
\noindent
It will be then just a matter of computation to prove the convergence of the sequence as defined by \eqref{mn}, \eqref{hn} to a limit pair.
\begin{prop}\label{p2.4}
Let $\left(m_n,h_n\right)$ recursively defined by \eqref{mn} and \eqref{hn}, with starting element $\left(m_0,h_0\right)$. For any $\ve<\overline{\ve}$ there is a couple $\left(m,h\right)$ in $C^{\infty}([0,\ve^{-1}])$ such that:
\begin{eqnarray}
&&\lim_{n\to\infty} \left\|m-m_n\right\|_{\infty} = 0, \\
&&\lim_{n\to\infty} \left\|h-h_n\right\|_{\infty} = 0
\end{eqnarray}
$m$ solving \eqref{pbs} with $m\left(0\right) = \mu^-_{\infty}$ for $x\in\left(-\infty,0\right)$ and $m\,(\ve^{-1}) = \mu^+_{\infty}$, and $h$ satisfying $h=T\left(m\right)$, where:
\begin{equation}
\mu^-_{\infty} \coloneqq \lim_{n\to\infty} \mu^-_n \qquad
\mu^+_{\infty} \coloneqq \lim_{n\to\infty} \mu^+_n
\end{equation}
and $(\mu^-_{\infty},\mu^+_{\infty})\in I_{\beta}\times I_{\beta}$.
\end{prop}
\begin{center}
\textit{Invertibility of the scheme}
\end{center}
The previous results actually imply that for any couple of boundary values $(\mu^-,\mu^+)$ and taking $m_0$ solution of \eqref{pb0} as starting element of the iteration, the sequence of couples converges to a smooth, bounded solution of \eqref{pbs} corresponding to the boundary conditions $(\mu^-_{\infty},\mu^+_{\infty})$, which in general differ of $O\left(\ve\right)$ from $(\mu^-,\mu^+)$. We thus need to prove that the map:
\begin{equation}
\left(\mu^-,\mu^+\right) \mapsto \left(\mu^-_{\infty},\mu^+_{\infty}\right)
\end{equation}
is one-to-one and then invertible. Indeed, this would imply that there always exists a certain starting element $(\mu^-,\mu^+)$ which is mapped to the desired couple of boundary magnetizations. 
\newline
We resume this result, which actually closes the proof of Theorem 2.1, in the following:
\begin{prop}\label{p2.5}
Let $\mathcal{F}:I_{\beta}\times I_{\beta}\mapsto A_{\beta}$, $A_{\beta}$ subset of $I_{\beta}\times I_{\beta}$, be the updating map of the boundary conditions:
\begin{equation}\label{map}
\mathcal{F}\left(\mu^-_n,\mu^+_n\right) = \left(\mu^-_{n+1},\mu^+_{n+1}\right)
\end{equation}
with $(\mu^-_0,\mu^+_0)\equiv(\mu_-,\mu_+)$. Then, for any $\ve<\overline{\ve}$:
\begin{equation}\label{jac}
\mathcal{J}_{\mathcal{F}} = \mathrm{diag}\,\left(1,1\right) + O\left(\ve\right)\cdot J_2
\end{equation}
where $\mathcal{J}_{\mathcal{F}}$ indicates the jacobian of \eqref{map}.
\end{prop}
\noindent
In \eqref{jac}, $J_2$ stands for the $2\times 2$ matrix in which all the entries are equal to one. This result says that, for $\ve$ small enough, $\mathrm{det}\, \mathcal{J}_{\mathcal{F}}>0$; in this case $\mathcal{F}$ is invertible and thus $A_{\beta}$ coincides with the open square $I_{\beta}$ when $\ve\to 0$. Proposition \ref{p2.5} actually closes the proof of Theorem \ref{theorem}.
\newline
\begin{center}
\textit{Formal approach}
\end{center}
The proof of Theorem 2.1 consists of two parts: firstly, we shall exhibit a feasible sequence of pairs $\left(m_n,h_n\right)$ which converges to a fixed point of \eqref{pbs} with certain boundary conditions; then, we have to prove that any couple $\left(\mu^-_{\infty},\mu^+_{\infty}\right)$ has a preimage $\left(\mu^-,\mu^+\right)$ in the sense of map \eqref{map}.
\newline
In the iteration, given $h_{n-1}$, we solve \eqref{mn} in order to get the new magnetization $m_n$ and, according to \eqref{hn}, we compute $h_n$ and so on. As clarified in \cite{DMPT}, in this recursion the main point consists in controlling the variation of the magnetization $\delta m$ when $h$ changes of an amount of $\delta h$. Without claiming to be complete here, consider the first variation of \eqref{pbs} with respect to the magnetic field:
\begin{equation}
\delta m\left(x\right) = \chi\left(m\left(x\right)\right) \Big[\left(J\ast \delta m\right)\left(x\right) + \delta h\left(x\right)\Big]
\end{equation}
namely:
\begin{equation}
\Big(I-\chi\left(m\left(x\right)\right)J\ast\,\Big) \delta m\left(x\right) = \chi\left(m\left(x\right)\right) \delta h\left(x\right)
\end{equation}
the term in brackets being a linear operator acting on $\delta m$ (see Section 3). If $\left\|\chi\left(m\right)\right\|_{\infty}<1$, $\mathscr{L}_{h,m}\coloneqq \chi\left(m\right) J \,\ast\,$ is invertible and, moreover, its 
inverse has an explicit form as a convergent series:
\begin{equation}\label{deltam}
\delta m\left(x\right) = \Big(\mathscr{I}-\mathscr{L}_{h,m}\Big)^{-1}\chi\left(m\left(x\right)\right) \delta h\left(x\right) = \sum_{k=0}^{\infty} \mathscr{L}_{h,m}^k \, \chi\left(m\left(x\right)\right) \delta h\left(x\right).
\end{equation}
Due to the presence of the convolution term, expression \eqref{deltam} makes clear that the pointwise change in the magnetization depends on the variation of $h$ on the whole real line. Nevertheless, this is not completely true since the interaction is long range but finite. This property of the convolution kernel is magnified and then exploited by the introduction of a weighted norm, which actually provides a way to bound $\delta m$.
\newline
One of the issues in \cite{DMPT} consists in the invertibility of $\mathscr{L}_{h,m}$, because an instanton-shape interface connects regions of different phases and, thus, a detailed analysis of the spectral properties of $\mathscr{L}_{h,m}$ is required, and we refer in particular to \cite{DMOP} for an exhaustive treatment of this.
In our case, the magnetization profile is entirely contained in $I_{\beta}$, so that $\left\|p_{h,m}\right\|_{\infty}<1$. What really matters here is the control of $\delta h$, together with its derivatives, with respect to the boundary magnetizations. 
\subsection{Structure of the work}

The paper is organized as follows:
\newline
\newline
in Section 3 we prove Proposition \ref{p2.2} by showing that there exists an underlying sequence which converges, in the sup norm, to the solution of \eqref{m1}. Uniform estimates are made for the magnetization profiles and the corresponding auxiliary magnetic fields.
\newline
\newline
Section 4 is devoted to the inductive proof of Proposition \ref{p2.3} and \ref{p2.4}. We carefully make certain that, in the iteration, the magnetization profile stays confined between $m^*\left(\beta\right)$ and $1$.
\newline
\newline
In Section 5 we collect the previous results and deal with the invertibility issue to finally prove Theorem 2.1.

\section{First iteration}

\begin{center}
\textit{Setting}
\end{center}
Following the notation introduced in \cite{EP}, we indicate the first derivative of the hyperbolic tangent with respect to the whole argument with:
\begin{equation}
p_{h,m}\left(x\right) \coloneqq {\beta \over  \cosh^2 \Big\{\beta\Big[\left(J\ast m\right)\left(x\right) + h\left(x\right)\Big]\Big\}}.
\end{equation}
Notice that $p_{h,m}\left(x\right)\equiv\chi\left(m\left(x\right)\right)$ provided $m\left(x\right)$ satisfies $m\left(x\right) = \tanh \Big\{\beta\Big[\left(J\ast m\right)\left(x\right) + h\left(x\right)\Big]\Big\}$. Consider then the Banach space $C^{\infty}([0,\ve^{-1}])$ equipped with the sup norm $\| \cdot \|_{\infty} = \| \cdot \|_{L^{\infty}{[0,\ve^{-1}]}}$ and define the main operator $\mathscr{L}_{h,m}\in \mathcal{L}\left(C^{\infty}([0,\ve^{-1}])\right)$, $\mathscr{L}_{h,m}: C^{\infty}([0,\ve^{-1}])\mapsto C^{\infty}([0,\ve^{-1}])$ through its action on a bounded function $f$:
\begin{equation}
\mathscr{L}_{h,m} f\left(x\right) \coloneqq \int_{\mathbb{R}} p_{h,m}\left(y\right) J\left(x,y\right) f\left(y\right) \text{d}y.
\end{equation}
If $\left\|p_{h,m}\right\|_{\infty}<1$, we are allowed to write:
\begin{equation}\label{sum}
\Big(\mathscr{I}-\mathscr{L}_{h,m}\Big)^{-1} = \sum_{k=0}^{\infty} \mathscr{L}^k_{h,m} \eqqcolon \mathscr{U}_{h,m}
\end{equation}
where $\mathscr{I}$ stands for the identity operator in the mentioned space. The action of the $k$-th power of $\mathscr{L}_{h,m}$ on $f\in C^{\infty}\left([0,\ve^{-1}]\right)$ is explicitly given by:
\begin{equation}
\mathscr{L}^k_{h,m} f\left(x_0\right) = \int_{\mathbb{R}^k} f\left(x_k\right) \prod_{j=1}^k p_{h,m}\left(x_j\right) J\left(x_{j-1},x_j\right) \text{d}x_j.
\end{equation}
We specify here that, in what follows, the norm of a linear operator acting on a $C^{\infty}([0,\ve^{-1}])$ function will be always the sup norm, i.e.:
\begin{equation}
\left\|\mathscr{L}_{h,m}\right\|_{\infty} \coloneqq \sup_{f\in C^{\infty}([0,\ve^{-1}])} 
{\left\|\mathscr{L}_{h,m} f \right\|_{\infty}\over \left\|f\right\|_{\infty}}.
\end{equation}
\newline
\begin{center}
\textit{The starting point}
\end{center}
By integrating \eqref{pb0} between $0$ and $\ve^{-1}$ we get the explicit expression of the mesoscopic current in terms of the boundary conditions:
\begin{equation}\label{j}
j =  \left(1-\beta\right)\left(\mu_- -\mu_+\right) + {\beta\over 3}\left(\mu_-^3-\mu_+^3\right) > 0.
\end{equation}
while the inverse of the unique real solution of \eqref{pb0}, depending on $\mu_-$ and $j$, is:
\begin{equation}\label{inv}
x =  \left(j\ve\right)^{-1}\left[\left(1-\beta\right)\left(\mu_- - m_0\left(x\right)\right) + {\beta\over 3}\left(\mu_- -m_0^3\left(x\right)\right)\right].
\end{equation}
Expression \eqref{inv} is actually invertible since:
\begin{equation}
\mathrm{sign}\, j = - \mathrm{sign} \,{\text{d}\over \text{d}x}m_0\left(x\right)
\end{equation}
furthermore, as a function of $m_0$, $x$ is smooth in $\left[\mu_-,\mu_+\right]$, and then $m_0\in C^{\infty}([0,\ve^{-1}])$. 
\newline
The choice of $\left(m_0,h_0\right)$ as the starting pair for the scheme is motivated by the fact that, as the size of the system increases, we expect the true stationary profile to get closer to $m_0\left(x\right)$ since, formally, the variation of the magnetization in a mesoscopic length is small, so that:
\begin{equation}
\int_{\mathbb{R}} J\left(x,y\right) {\text{d}\over \text{d}y} m\left(y\right) \text{d}y \simeq \int_{\mathbb{R}} J\left(x,y\right) {\text{d}\over \text{d}x} m\left(x\right) \text{d}y = {\text{d}\over \text{d}x} m\left(x\right)
\end{equation}
thus, equation \eqref{pb} should reduce to \eqref{pb0} as $\ve\to 0$.

\subsection{Proof of Proposition \ref{p2.2}}

\begin{center}
\textit{Working hypothesis}
\end{center}
The purpose is to write the solution of \eqref{m1} as an infinite sum:
\begin{equation}\label{m1sum}
m_1\left(x\right) = m_0\left(x\right) + \sum_{k=1}^{\infty} \vf_{1,k}\left(x\right)
\end{equation}
where each $\vf_{1,k}$ depends on $m_0$ and on the former $\vf_{1,j}$'s, with $\vf_{1,k}\in C^{\infty}([0,\ve^{-1}])$ for any $k\in\mathbb{N}^*$. As we shall see, the scheme actually reduces to a functional Newton's interpolation method. 
\newline
By hypothesis, there exists $\delta>0$ such that $\mu_+>m^*\left(\beta\right)+\delta$ and $\mu_-<1-\delta$. We now fix $0<\delta'<\delta$ and construct the series \eqref{m1sum} in such a way that:
\begin{equation}
m^*\left(\beta\right)+\delta' < \left\|m_0+\phi_{1,k}\right\|_{\infty}< 1-\delta'
\end{equation}
where, for notational convenience:
\begin{equation}
\phi_{1,k}\left(x\right) \coloneqq \sum_{j=1}^{k} \vf_{1,j}\left(x\right) \qquad k\in\mathbb{N}^*.
\end{equation}
In this hypothesis, there is a constant $\lambda\left(\delta'\right)$ such that:
\begin{equation}
\left\| p_{h_0,m_0+\phi_{1,k}} \right\|_{\infty} < \lambda\left(\delta'\right) < 1
\end{equation}
for any $k\in\mathbb{N}^*$. We give an explicit expression for $\lambda$ observing that:
\begin{eqnarray}
\left\| J\ast\left(m_{0}+\phi_{1,k}\right) + h_0\right\|_{\infty} &\le& \max\Big\{
\left|1-\delta'-{j\chi^{-1}\left(m^*\left(\beta\right)+\delta'\right)}\right|,\left|m^*\left(\beta\right)+\delta'- j \chi^{-1}\left(1-\delta'\right)\right|\Big\} \nn \\
&\eqqcolon& \zeta_{\delta'}
\end{eqnarray}
so that, as a consequence:
\begin{equation}
\left\| p_{h_0,m_0+\phi_{1,k}} \right\|_{\infty} < \lambda\left(\delta'\right) \eqqcolon{\beta\over\cosh^2\left(\beta\zeta_{\delta'}\right)}
\qquad
\left\| \mathscr{U}_{h_0,m_0+\phi_{1,k}} \right\|_{\infty} < {1\over 1-\lambda\left(\delta'\right)} \eqqcolon \overline{u}\left(\delta'\right).
\end{equation}
In the following sections, the dependence on $\delta'$ is dropped to lighten the notation.
\newline
\begin{center}
\textit{Sequence of iterates}
\end{center}
In this hypothesis define:
\begin{eqnarray}
\vf_{1,1}\left(x\right) &\coloneqq& \mathscr{U}_{h_0,m_0} \left(\tanh\Big\{\beta\Big[\left(J\ast m_0\right)\left(x\right) + h_{0}\left(x\right)\Big]\Big\} - m_0\left(x\right)\right) \label{phi1} \\
\vf_{1,{k+1}}\left(x\right) &\coloneqq&\mathscr{U}_{h_0,m_0+\phi_{1,k}}\Big(\tanh\Big\{\beta\Big[\Big(J\ast \left(m_0+\phi_{1,k}\right)\Big)\left(x\right) + h_{0}\left(x\right)\Big]\Big\} 
- m_0\left(x\right) - \phi_{1,k}\left(x\right)\Big). \label{phin} \hspace{1cm}
\end{eqnarray}
The following result holds true:
\begin{prop} For any $\ve<\ve^*$, where:
\begin{equation}
\ve^*\coloneqq\min\Big\{{1\over 2\,\overline{u}^2c^2},{\delta-\delta'\over 2\,\overline{u}c} \Big\},
\end{equation}
there are constants $0<a,b<\infty$ such that
\begin{flalign}
(\mathrm{a})& & \left\|\vf_{1,1}\right\|_{\infty} \le& \, \overline{u}  a \ve \label{it1}  \\
(\mathrm{b})& & \left\|\vf_{1,k+1}\right\|_{\infty} \le& \, \overline{u} b \left\|\vf_{1,k}\right\|^2_{\infty} \label{itk} &
\end{flalign}
for any $k\in\mathbb{N}^*$.
\end{prop}
\noindent
\textbf{Proof.} (a)
\newline
Compute:
\begin{eqnarray}
\left|\tanh\Big\{\beta\Big[\left(J\ast m_0\right)\left(x\right) + h_{0}\left(x\right)\Big]\Big\} - m_0\left(x\right) \right| &\le& \left|  \tanh\Big\{\beta\Big[\left(J\ast m_0\right)\left(x\right) - m_{0}\left(x\right)\Big]\Big\}\right| \nn \\
&\le& \beta \left| \left(J\ast m_0\right) - m_0\left(x\right)\right| \nn \\
&\le& \beta \sup_{y\in[x-1,x+1]} \left|m_0\left(y\right)-m_0\left(x\right)\right|. 
\end{eqnarray}
Using \eqref{pb0} we get:
\begin{equation}
\sup_{y\in(x,x+1]} \left|m_0\left(y\right)-m_0\left(x\right)\right| \le \sup_{y\in(x,x+1]}
\left|{\text{d}m_0\over \text{d}y}\left(y\right)\right| < {j \over 1-\beta\left(1-(\mu^+)^2\right)} \ve
\end{equation}
hence:
\begin{equation}\label{336}
\left\| \tanh\Big\{\beta\Big[\left(J\ast m_0\right) + h_{0}\Big]\Big\}- m_0 \right\|_{\infty} \le {\beta j\over 1-\beta\left(1-(\mu^+)^2\right)} \ve \eqqcolon a \ve
\end{equation}
and then, by definition of $\vf_{1,1}$, the result.
\qed
\newline
\newline
\textbf{Proof.} (b)
Suppose \eqref{itk} to be true for any $j\le k$ so that $\vf_{k+1}$ is actually well defined. Expand in Taylor series the hyperbolic tangent at the $k+1$-th step as follows:
\begin{eqnarray}
\tanh\Big\{\beta\Big[\Big(J\ast \left(m_0+\phi_{1,k}\right)\Big)\left(x\right) + h_{0}\left(x\right)\Big]\Big\} &=& \tanh\Big\{\beta\Big[\Big(J\ast \left(m_0+\phi_{1,k-1}\right)\Big)\left(x\right) + h_{0}\left(x\right)\Big]\Big\} \nn \\
&+& p_{h_0,m_0+\phi_{1,k-1}}\left(x\right)\left(J\ast \vf_{1,k}\right)\left(x\right) \nn \\
&-& p'_{h_0,m_0+\phi_{1,k-1}}\left(x\right)\left(J\ast \vf_{1,k}\right)^2\left(x\right) \nn \\
&+& O\left(\left(J\ast \vf_{1,k}\right)^3\left(x\right)\right)
\end{eqnarray}
where:
\begin{equation}\label{expan}
p'_{h,m}\left(x\right) \coloneqq p_{h,m}\left(x\right) \tanh\Big\{\beta\Big[\left(J\ast m\right)\left(x\right) + h\left(x\right)\Big]\Big\}.
\end{equation}
Combining the definition of $\vf_{1,k}$ with \eqref{expan} we get:
\begin{equation}
\vf_{1,k+1}\left(x\right) = - \mathscr{U}_{h_0,m_0+\phi_{1,k}} \Big(p'_{h_0,m_0+\phi_{1,k-1}}\left(x\right)\left(J\ast \vf_{1,k}\right)^2\left(x\right) + O\left(\left(J\ast \vf_{1,k}\right)^3\left(x\right)\right) \Big)
\end{equation}
which implies:
\begin{equation}
\left\|\vf_{1,k+1}\right\|_{\infty} \le \overline{u} \sup_{x'\in[0,\ve^{-1}]}p'_{h_0,m_0+\phi_{1,k-1}}(x') \sup_{x''\in[0,\ve^{-1}]} \left(J\ast \vf_{1,k}\right)^2(x'').
\end{equation}
The terms in the right brackets are uniformly bounded, therefore:
\begin{equation}
\left\|\vf_{1,k+1}\right\|_{\infty} \le \, \overline{u}b \left\|\vf_{1,k}\right\|^2_{\infty}
\end{equation}
provided:
\begin{equation}
b \ge \sup_{x'\in[0,\ve^{-1}]}p'_{h_0,m_0+\phi_{1,k-1}}(x')
\end{equation}
where $b$ do not depend on $k$.
\qed
\begin{center}
\textit{Convergence of the first term}
\end{center}
By iteration of \eqref{itk}, for any $k\in\mathbb{N}^*$:
\begin{equation}
\left\|\vf_{1,k+1}\right\|_{\infty} \le \left(\overline{u}^2 b\right)^{2^k-1} \left\|\vf_{1,1}\right\|^{2^k}_{\infty}
\end{equation}
so that, denoted with $c$ the maximum among $a$ and $b$, by \eqref{it1}:
\begin{equation}
\left\|\vf_{1,k+1}\right\|_{\infty} \le {1\over\overline{u}c}\left(\overline{u}^2 c^2\right)^{2^k} \ve^{2^k};
\end{equation}
summing on $k$:
\begin{equation}\label{sconv}
\sum_{k=1}^{\infty} \left\|\vf_{1,k}\left(x\right)\right\|_{\infty} \le \overline{u}c\ve + {1\over\overline{u}c}\sum_{k=1}^{\infty} \left(\overline{u}^2 c^2 \ve\right)^{2^k}.
\end{equation}
Being $\ve<1/2\overline{u}^2 c^2$, we get form \eqref{sconv}:
\begin{equation}\label{347}
\left\|m_1-m_0\right\|_{\infty} \le \sum_{k=1}^{\infty} \left\|\vf_{1,k}\right\|_{\infty} \le  2\,\overline{u}c\ve < \delta-\delta'.
\end{equation}
The uniform convergence to the solution $m_1$ of \eqref{m1} directly follows from the definition of the $\vf_{1,j}$'s (Newton's method).
\qed

\section{Convergence to the mesoscopic profile}

\begin{center}
\textit{Notation and preliminaries}
\end{center}
\noindent
In the previous section we chose in a feasible way the starting point for the iteration and computed the new element $\left(m_1,h_1\right)$, with $m_1\in I_{\beta}$ for any $x\in[0,\ve^{-1}]$ and solving the mean field-type problem \eqref{m1}. We shall repeat this scheme to compute a new pair $\left(m_2,h_2\right)$, satisfying similar equations, and so on, acting in such a way that $\left(m_n,h_n\right)$ converges to the solution of \eqref{pbs} when $n\to\infty$. We expect that this solution, in turn, converges to the macroscopic profile when $\ve$ goes to zero. 
\newline
In this section we prove, by induction, that this convergence is achieved in the weighted $\alpha$-norm:
\begin{equation}
\left\| f \right\|_{\alpha} \coloneqq \sup_{x\in[0,\ve^{-1}]} \mathrm{e}^{-\alpha\ve x} \left|f\left(x\right)\right|\qquad f\in L^{\infty}\left([0,\ve^{-1}]\right)
\end{equation}
provided $\alpha$ is large enough and $\ve$ suitably small. At the same time, we control the uniform variation with respect to $m_0$ to ensure that at each iteration the magnetization does not enter the forbidden region and, thus, $\left\|p_{h_n,m_m}\right\|_{\infty}<1$. 
\newline
The whole strategy consists in finding sequences of $C^{\infty}([0,\ve^{-1}])$ functions $\left(\vf_{n,k}\left(x\right)\right)_{k=1}^{\infty}$ such that:
\begin{equation}\label{phinpo}
m_{n}\left(x\right) = m_{n-1}\left(x\right) + \sum_{k=1}^{\infty} \vf_{n,k}\left(x\right) \qquad \forall n\in\mathbb{N}^*
\end{equation}
where $m_n$ solves equation \eqref{mn} with certain boundary conditions. 
\newline
Before proving Proposition \ref{p2.3} define, in a consistent way with \eqref{phi1}, \eqref{phin}:
\begin{eqnarray}
\vf_{n,1}\left(x\right) &=& \mathscr{U}_{h_{n-1},m_{n-1}} \left(\tanh\Big\{\beta\Big[\left(J\ast m_n\right)\left(x\right) + h_{n}\left(x\right)\Big]\Big\} - m_n\left(x\right)\right) \label{phi1n} \\ 
\vf_{n,k+1}\left(x\right) &=&\mathscr{U}_{h_{n-1},m_{n-1}+\phi_{n,k}}\Big(\tanh\Big\{\beta\Big[\Big(J\ast \left(m_n+\phi_{n,k}\right)\Big)\left(x\right) + h_{n}\left(x\right)\Big]\Big\} 
- m_n\left(x\right) - \phi_{n,k}\left(x\right)\Big) \label{nk1}\nn \\
\end{eqnarray}
being:
\begin{equation}
\phi_{n,k}\left(x\right) \coloneqq \sum_{j=1}^{k} \vf_{n,j}\left(x\right) \qquad k\in\mathbb{N}^*.
\end{equation}
Again, we work in the hypothesis that, in this construction:
\begin{equation}
m^*\left(\beta\right)+\delta' < \left\|m_n+\phi_{n,k}\right\|_{\infty}< 1-\delta'
\end{equation}
so that \eqref{phi1n}, \eqref{nk1} actually exist and $\left\| p_{h_n,m_n+\phi_{n+1,k}} \right\|_{\infty} < \lambda < 1$.

\subsection{Proof of Proposition \ref{p2.3}}

We prove that for any choice of $\alpha$ and $\ve$ such that:
\begin{eqnarray}
\alpha &>& {8j\overline{u} \over \delta'\left(2-\delta'\right)} \label{alphadef} \\
\ve &<& \min\left\{{1\over\alpha}\log{1+\lambda\over\lambda},{\mathrm{e}^{2\alpha}\over 2}\ve^*\right\}\eqqcolon \widetilde{\ve} \label{epsdef}
\end{eqnarray}
the sequence of $\left(m_n,h_n\right)$ is well defined for any $n\in\mathbb{N}$ and solves equations \eqref{mn}, \eqref{hn}. 
\newline
Fix $n\in\mathbb{N}^*$ and suppose that to be true for any $j\le n$ so that similar bounds to those provided for the first iterate hold true. Compute, for any $x\in[0,\ve^{-1}]$, the magnetic field variation:
\begin{eqnarray}
\left| h_{n+1}\left(x\right) - h_n\left(x\right)\right| &\le& {j\over \beta\delta'\left(2-\delta'\right)}\ve\int_0^x \left| m_{n+1}^2\left(y\right) - m_n^2\left(y\right)\right| \text{d}y \nn \\
&\le& {2j\over \beta\delta'\left(2-\delta'\right)}\ve\int_0^x \left| m_{n+1}\left(y\right) - m_{n}\left(y\right)\right| \text{d}y \nn \\
&\le& {C\over\beta}\ve\, \left\| m_{n+1}-m_n\right\|_{\alpha} \int_0^x \mathrm{e}^{\alpha\ve y} \, \text{d}y \nn \\
&\le& {C\over\beta\alpha} \, {\mathrm{e}^{\alpha\ve x}}
\, \left\| m_{n+1}-m_n\right\|_{\alpha} 
\end{eqnarray}
so that:
\begin{equation}
\left| h_{n+1}\left(x\right) - h_n\left(x\right)\right|{\mathrm{e}^{-\alpha\ve x}}  \le {C\over\beta\alpha} \left\| m_{n+1}-m_n\right\|_{\alpha}.
\end{equation}
Taking the supremum with respect to $x$ we get the inequality in the $\alpha$-norm:
\begin{equation}\label{itera}
\left\| h_{n+1}-h_n\right\|_{\alpha} \le {C\over\beta\alpha}\left\| m_{n+1}-m_n\right\|_{\alpha}.
\end{equation}
A preliminary result is needed at this point.
\begin{lem}\label{lemf}
Let $f,g\in C^{\infty}\left([0,\ve^{-1}]\right)$ with:
\begin{equation}
\left|f\left(x_0\right)\right| \le \mathscr{U}_{h,m} \left|g\left(x_0\right)\right| \qquad \forall x_0\in[0,\ve^{-1}].
\end{equation}
Then, at fixed $\alpha>0$:
\begin{equation}
\left\| f\right\|_{\alpha} \le {\left\| g\right\|_{\alpha}\over 1-\lambda\mathrm{e}^{\alpha\ve}}
\end{equation}
provided $\ve<{1\over\alpha}\log\lambda$.
\end{lem}
\noindent
{\bf Proof.}
We have:
\begin{eqnarray}
\mathscr{L}^k_{h,m} \left|g\left(x_0\right)\right| &=& \int_{\mathbb{R}^k} \left|g\left(x_k\right)\right| \prod_{j=1}^k p_{h,m}\left(x_j\right) J\left(x_{j-1},x_j\right) \text{d}x_j \nn \\
&\le& \lambda^k \, \left\| g\right\|_{\alpha}\int_{\mathbb{R}^k} \mathrm{e}^{\alpha\ve x_k}\prod_{j=1}^k J\left(x_{j-1},x_j\right) \text{d}x_j 
\end{eqnarray}
hence:
\begin{equation}
\left|f\left(x_0\right)\right| \le \left\| g\right\|_{\alpha}\sum_{k=0}^{\infty} \lambda^k \, \int_{\mathbb{R}^k} \mathrm{e}^{\alpha\ve x_k}\prod_{j=1}^k J\left(x_{j-1},x_j\right) \text{d}x_j.
\end{equation}
Multiplying both the sides by $\mathrm{e}^{-\alpha \ve x_0}$ we get:
\begin{eqnarray}
\left|f\left(x_0\right)\right|\mathrm{e}^{-\alpha \ve x_0} &\le& \left\| g\right\|_{\alpha}\sum_{k=0}^{\infty} \lambda^k \, \int_{\mathbb{R}^k} \mathrm{e}^{\alpha\ve \left|x_k-x_0\right|} \prod_{j=1}^k J\left(x_{j-1},x_j\right) \text{d}x_j \nn\\
&\le& \left\| g\right\|_{\alpha}\sum_{k=0}^{\infty} \left(\lambda\mathrm{e}^{\alpha\ve}\right)^k =  {\left\| g\right\|_{\alpha}\over 1-\lambda\mathrm{e}^{\alpha\ve}}.
\end{eqnarray}
Taking the supremum with respect to $x_0$, we get the result.
\qed
\newline
\newline
Consider now the first correction to the $n$-th iterate:
\begin{eqnarray}
\left|\vf_{n+1,1}\left(x\right)\right| &\le& \mathscr{U}_{h_n,m_n} \left| \tanh\Big\{\beta\Big[\left(J\ast m_{n}\right)\left(x\right) + h_{n}\left(x\right)\Big]\Big\} -m_n\left(x\right)\right| \nn \\
&\le& \beta \mathscr{U}_{h_n,m_n} \left| h_n\left(x\right) -h_{n-1}\left(x\right)\right|;
\end{eqnarray}
given the previous bound:
\begin{equation}
\left\|\vf_{n+1,1}\right\|_{\alpha} \le {\beta\over 1-\lambda\mathrm{e}^{\alpha\ve}} \left\| h_{n}-h_{n-1} \right\|_{\alpha} \le {1\over 1-\lambda\mathrm{e}^{\alpha\ve}} {C\over\alpha} \left\| m_{n}-m_{n-1} \right\|_{\alpha}
\end{equation}
and, applying again Lemma \ref{lemf} to the $k+1$-th correction \eqref{nk1}:
\begin{equation}
\left\|\vf_{n+1,k+1}\right\|_{\alpha} \le \overline{u}c \left\|\vf_{n+1,k}\right\|^2_{\alpha}.
\end{equation}
where the last inequality follows from the assumptions \eqref{alphadef}, \eqref{epsdef}.
\newline
Summing on $k$ we get:
\begin{equation}
\sum_{k=1}^{\infty} \left\|\vf_{n+1,k}\right\|_{\alpha} \le \left\|\vf_{n+1,1}\right\|_{\alpha} + {1\over\overline{u}c} \sum_{k=1}^{\infty} \left(\overline{u}c \left\|\vf_{n+1,1}\right\|_{\alpha}\right)^{2^k}
\end{equation}
and then:
\begin{equation}
\left\| m_{n+1}-m_{n} \right\|_{\alpha} \le 2 \left\|\vf_{n+1,k}\right\|_{\alpha}\le {2\over 1-\lambda\mathrm{e}^{\alpha\ve}} {C\over\alpha} \left\| m_{n}-m_{n-1} \right\|_{\alpha} \le {1\over 2}\left\| m_{n}-m_{n-1} \right\|_{\alpha}.
\end{equation}
Therefore, by recursion:
\begin{equation}\label{contr}
\left\| m_{n+1}-m_{n} \right\|_{\alpha} \le \left({1\over 2}\right)^n\left\|m_{1}-m_{0} \right\|_{\alpha}
\le 2\left({1\over 2}\right)^n \overline{u}c\ve.
\end{equation}
\eqref{contr} is a contraction in the $\alpha$-norm. Furthermore, we can immediately check that at each iteration the magnetization profile is still in $I_{\beta}$. Indeed, it has to be:
\begin{equation}
\left\| m_{n}\left(x\right) - m_0\left(x\right)\right\|_{\infty} \le \delta-\delta' \qquad \forall n\in\mathbb{N}
\end{equation}
but:
\begin{eqnarray}
\left\| m_{n}\left(x\right) - m_0\left(x\right)\right\|_{\infty} &\le& \sum_{k=0}^{n} \left\| m_{k+1}\left(x\right) - m_k\left(x\right)\right\|_{\infty} \le \sum_{k=0}^{\infty} \left\| m_{k+1}\left(x\right) - m_k\left(x\right)\right\|_{\infty} \nn \\
&\le&2\mathrm{e}^{2\alpha}\,\overline{u}c\ve \sum_{k=0}^{\infty} \left({1\over 2}\right)^k 
\le 4\mathrm{e}^{2\alpha}\,\overline{u}c\ve
\end{eqnarray}
where we used the fact that $\left\|\,\cdot\,\right\|_{\alpha} \le \mathrm{e}^{2\alpha}\left\|\,\cdot\,\right\|_{\infty}$. 
\newline
Clearly, a similar recurrence relation holds for the variation of subsequent magnetic fields.
\qed

\subsection{Proof of Proposition \ref{p2.4}}

As a consequence of Proposition \ref{p2.3}, there exists a limit pair $\left(m,h\right)$, with $m,h\in C^{\infty}([0,\ve^{-1}])$ such that:
\begin{eqnarray}
\lim_{n\to\infty} \left\|m_n-m\right\|_{\alpha} = 0 \\
\lim_{n\to\infty} \left\|h_n-h\right\|_{\alpha} = 0.
\end{eqnarray}
By continuity:
\begin{equation}
\tanh\Big\{\beta\Big[\left(J\ast m_{n}\right)\left(x\right) + h_{n}\left(x\right)\Big]\Big\} \xrightarrow[n\to\infty]{\alpha\text{-norm}}\tanh\Big\{\beta\Big[\left(J\ast m\right)\left(x\right) + h\left(x\right)\Big]\Big\} 
\end{equation}
thus, $\left(m_n,h_n\right)$ actually converges to the solution of \eqref{pb} with certain boundary magnetizations $(\mu^-_{\infty},\mu^+_{\infty})$ when $n$ goes to infinity.
\qed

\section{Invertibility}

Given a stable or metastable couple $(\mu^-,\mu^+)$, we proved so far that the iterations lead to a new pair of boundary magnetizations for which \eqref{pb} admits a smooth bounded solution, provided $\ve$ is small enough. However, since we are interested in finding a profile satisfying \eqref{pb} with prescribed conditions, we shall prove that the mentioned scheme is invertible, that is \eqref{map} is one-to-one; in this case, we could suitably fix the initial magnetizations which are mapped to the established ones. Nevertheless, this is not trivial at all since this dependence is deeply implicit.
\newline
\begin{center}
\textit{Macroscopic boundary variations}
\end{center}
Consider the expression \eqref{j} of the mesoscopic current and derive it with respect to $\mu^-$ and $\mu^+$:
\begin{eqnarray}
\partial_{\mu^-}j &=& \Big(1-\chi\,(\mu^-)\Big)\ve \\
\partial_{\mu^+}j &=& -\Big(1-\chi\,(\mu^+)\Big)\ve \label{j+}
\end{eqnarray}
so that, from \eqref{inv}:
\begin{eqnarray}
\partial_{\mu^-}m_0\left(x\right) &=& {1-\chi\,(\mu^-) \over 1-\chi\left(m_0\left(x\right)\right)} \left(1-\ve x\right) \label{dm0-} \\
\partial_{\mu^+}m_0\left(x\right) &=& {1-\chi\,(\mu^+) \over 1-\chi\left(m_0\left(x\right)\right)}\ve x  \label{dm0+}
\end{eqnarray}
and, as expected:
\begin{eqnarray}
\Big(\partial_{\mu^-}m_0\left(0\right),\partial_{\mu^+}m_0\left(0\right) \Big) &=& \left(1,0\right) \\
\Big(\partial_{\mu^-}m_0\,(\ve^{-1}),\partial_{\mu^+}m_0\,(\ve^{-1})\Big) &=& \left(0,1\right).
\end{eqnarray}
It is worth writing the jacobian of $\mathcal{F}$ as $\mathcal{J}_{\mathcal{F}} = \mathcal{I} + \Delta$ where, explicitly:
\begin{equation}
\Delta \coloneqq
\begin{pmatrix}
\partial_{\mu^-}m\left(0\right) - \partial_{\mu^-}m_0\left(0\right)& \partial_{\mu^-}m\,(\ve^{-1}) - \partial_{\mu^-}m_0\,(\ve^{-1}) \\[4ex]
\partial_{\mu^+}m\left(0\right) -\partial_{\mu^+}m_0\left(0\right) & \partial_{\mu^+}m\,(\ve^{-1}) - \partial_{\mu^+}m_0\,(\ve^{-1})
\end{pmatrix}
\end{equation}
indeed, if we are able to prove that $\mathrm{det}\, \Delta = O\left(1\right)\cdot\ve$, there will exists $\ve^*$ such that for any $\ve<\ve^*$ the determinant of the jacobian of \eqref{map} is strictly positive and, thus, \eqref{map} is invertible. 

\subsection{Proof of Proposition \ref{p2.5}}

In the sequel, $\mu$ indicates both the left and the right boundary magnetizations.
\newline
\begin{center}
\textit{Uniform bound on the first iterate}
\end{center}
\begin{prop} 
There is $d>0$ such that:
\begin{equation}
\left \|\partial_{\mu}\vf_{1,1}\right\|_{\infty} \le d\ve.
\end{equation}
\end{prop}
\noindent
{\bf Proof.} 
Derive \eqref{phi1} and \eqref{noj} with respect to $\mu$:
\begin{eqnarray}
\partial_{\mu}\vf_{1,1}\left(x\right) &=& \mathscr{U}_{h_0,m_0}\Big\{
p_{h_0,m_0}\left(x\right)\left(J\ast \partial_{\mu}m_0\right)\left(x\right) + p_{h_0,m_0}\left(x\right) \partial_{\mu}h_0\left(x\right) - \partial_{\mu}m_0\left(x\right) \nn \\
&+& \partial_{\mu}p_{h_0,m_0}\left(x\right)\left(J\ast \vf_{1,1}\right)\left(x\right)\Big\} \\
\partial_{\mu}m_0\left(x\right) &=& \chi\left(m_0\left(x\right)\right) \Big[
\partial_{\mu}m_0\left(x\right) - \partial_{\mu}h_0\left(x\right)\Big].
\end{eqnarray}
Combining the previous expressions we get:
\begin{eqnarray}
\partial_{\mu}\vf_{1,1}\left(x\right)&=& \mathscr{U}_{h_0,m_0}\Big\{
p_{h_0,m_0}\left(x\right)\left(J\ast \partial_{\mu}m_0\right)\left(x\right) - \chi\left(m_0\left(x\right)\right)\partial_{\mu}m_0\left(x\right) \nn \\
&+& \Big[p_{h_0,m_0}\left(x\right)-\chi\left(m_0\left(x\right)\right)\Big]\partial_{\mu}h_0\left(x\right) +\partial_{\mu}p_{h_0,m_0}\left(x\right)\left(J\ast \vf_{1,1}\right)\left(x\right) \Big\}. \nn \\
\end{eqnarray}
Add and subtract $p_{h_0,m_0}\partial_{\mu}h_0$ to estimate:
\begin{eqnarray}
\partial_{\mu}\vf_{1,1}\left(x\right) &\le& \overline{u} \, \Big\{
p_{h_0,m_0}\left(x\right) \Big[\left(J\ast \partial_{\mu}m_0\right)\left(x\right) - \partial_{\mu}m_0\left(x\right)\Big] \nn \\
&+& \Big[p_{h_0,m_0}\left(x\right)-\chi\left(m_0\left(x\right)\right)\Big] \Big[\partial_{\mu}m_0\left(x\right)+\partial_{\mu}h_0\left(x\right)\Big] \nn \\
&+& \partial_{\mu}p_{h_0,m_0}\left(x\right)\left(J\ast \vf_{1,1}\right)\left(x\right) \Big\}
\end{eqnarray}
and notice that the following estimates hold:
\begin{eqnarray}
\left\|\left(J\ast \partial_{\mu}m_0\right) - \partial_{\mu}m_0 \right\|_{\infty} &\le& \ve
\\
\left\| p_{h_0,m_0}-\chi\left(m_0\right)\right\|_{\infty} &\le& 2\beta \left\| \tanh\Big\{\beta\Big[\left(J\ast m_{0}\right)+ h_{0}\Big]\Big\} - m_0 \right\|_{\infty} \le 2\beta c\ve.
\\
\left\|\partial_{\mu}p_{h_0,m_0}\right\|_{\infty} &\le& 2\beta\lambda \left\|\left(J\ast \partial_{\mu}m_0\right) + \partial_{\mu}h_0 \right\|_{\infty} \le 2\beta\lambda \left[b'+j\chi^{-1}(\mu^-)+\ve\right]. \,\,\,\,\,\,
\end{eqnarray}
where:
\begin{equation}
b'\coloneqq \left(1+j\right)\chi^{-1}(\mu^-) -1.
\end{equation}
Therefore:
\begin{equation}
\left \| \partial_{\mu}\vf_{1,1}\right\|_{\infty} \le
\left(\lambda+4\beta b'c +4\beta \overline{u}c \ve\right)\overline{u}\ve
\end{equation}
Choose then, e.g.:
\begin{equation}
d\coloneqq 2\max \left\{\lambda,4\beta\overline{u}b'c \right\}
\end{equation}
to get the result.
\qed
\begin{prop}
There is a constant $\gamma>0$ such that for any $k\in\mathbb{N}^*$:
\begin{equation}
\left \|\partial_{\mu}\vf_{1,k+1}\right\|_{\infty} \le \gamma^{2^{k+1}} \ve^{2^k}.
\end{equation}
\end{prop}
\noindent
{\bf Proof.}
The derivative of $\vf_{1,k+1}$ with respect to $\mu$ consists of two contributes.
\newline
\newline
\underline{\it First term}
\begin{equation}
\Big(\partial_{\mu}\mathscr{U}_{h_0,m_0+\phi_{1,k}}\Big)\Big(\tanh\Big\{\beta\Big[\Big(J\ast \left(m_0+\phi_{1,k}\right)\Big)\left(x\right) + h_{0}\left(x\right)\Big]\Big\} - m_0\left(x\right)- \phi_{1,k}\left(x\right)\Big) 
\end{equation}
\underline{\it Second term}
\begin{equation}
\mathscr{U}_{h_0,m_0+\phi_{1,k}}\partial_{\mu}\Big(\tanh\Big\{\beta\Big[\Big(J\ast \left(m_0+\phi_{1,k}\right)\Big)\left(x\right) + h_{0}\left(x\right)\Big]\Big\} - m_0\left(x\right)- \phi_{1,k}\left(x\right)\Big) 
\end{equation}
The first term has the same order as $\vf_{1,k+1}$; for what concerns the second one, we write the contribute in brackets as the rest of order two which remains form the Taylor functional expansion of the hyperbolic tangent. Since its second G\^ateaux derivative is:
\begin{equation}
\text{d}^2 \tanh\left(u;v\right) = -2 v^2 \tanh\left(u\right)\left[1-\tanh^2\left(u\right)\right]
\end{equation}
in the proper regolarity hypothesis on $u$ and $v$, once defined:
\begin{eqnarray}
u_k\left(x\right) &\coloneqq& \beta\Big[\Big(J\ast \left(m_0+\phi_{1,k-1}\right)\Big)\left(x\right) + h_0\left(x\right)\Big] \\
\Delta u_k\left(x\right) &\coloneqq& \beta \left(J\ast \vf_{1,k}\right)\left(x\right)
\end{eqnarray}
we get:
\begin{eqnarray}
R^{\,(2)}_{u_k,\Delta u_k} = -2 \int_0^1 \left(1-t\right) \tanh\left(u_k+t\Delta u_k\right)\left[1-\tanh^2\left(u_k+t\Delta u_k\right)\right] \Delta u^2_k \, \text{d}t.
\end{eqnarray}
The derivative with respect to $\mu$ is thus given by:
\begin{eqnarray}
\partial_{\mu}R^{\,(2)}_{u_k,\Delta u_k} &=& -2 \int_0^1 \left(1-t\right) 
\left[1-\tanh^2\left(u_k+t\Delta u_k\right)\right]^2 \Delta u^2_k \left(\partial_{\mu}u_k + t\partial_{\mu}\Delta u_k \right) \, \text{d}t \nn \\
&+& 4 \int_0^1 \left(1-t\right) \tanh^2\left(u_k+t\Delta u_k\right)\left[1-\tanh^2\left(u_k+t\Delta u_k\right)\right] \Delta u^2_k\left(\partial_{\mu}u_k + t\partial_{\mu}\Delta u_k \right) \, \text{d}t \nn \\
&-& 4 \int_0^1 \left(1-t\right) \tanh^2\left(u_k+t\Delta u_k\right)\left[1-\tanh^2\left(u_k+t\Delta u_k\right)\right]
\Delta u_k \partial_{\mu}\Delta u_k \, \text{d}t
\end{eqnarray}
from which the estimate in the sup norm:
\begin{equation}
\left\|\partial_{\mu}R^{\,(2)}_{u_k,\Delta u_k}\right\|_{\infty} \le
\lambda\left(2\left\|\Delta u_k\right\|_{\infty} + \left\|\Delta u_k\right\|_{\infty}^2
\right)\left\|\partial_{\mu}\Delta u_k\right\|_{\infty} 
+ 3\lambda \left\|\Delta u_k\right\|_{\infty}^2\left\|\partial_{\mu}u_k\right\|_{\infty}.
\end{equation}
Since:
\begin{eqnarray}
\left\|\partial_{\mu}u_k\right\|_{\infty} &\le&\beta \left\|\partial_{\mu}m_0\right\|_{\infty} +\beta \sum_{j=1}^{k-1} \left\|\partial_{\mu}\vf_{1,j}\right\|_{\infty} + \beta \left\|\partial_{\mu}h_0\right\|_{\infty} \\
\left\|\partial_{\mu}\Delta u_k\right\|_{\infty} &\le& \beta \left\|\partial_{\mu}\vf_{1,k}\right\|_{\infty}
\end{eqnarray}
we have:
\begin{equation}\label{resto}
\left\|\partial_{\mu}R^{\,(2)}_{u_k,\Delta u_k}\right\|_{\infty} \le 
3\beta^2 \lambda \Bigg[\Bigg(\sum_{j=1}^{k} \left\|\partial_{\mu}\vf_{1,j}\right\|_{\infty} + \chi^{-1}\left(\mu^-\right)\Bigg)\left\|\vf_{1,k}\right\|_{\infty}^2 + \left\|\partial_{\mu}\vf_{1,k}\right\|_{\infty}\left\|\vf_{1,k}\right\|_{\infty}\Bigg]
\end{equation}
hence, combining the two terms, the uniform bound:
\begin{equation}\label{est}
\left\|\partial_{\mu}\vf_{1,k+1}\right\|_{\infty} \le 3\beta^2 \lambda\overline{u} \Bigg[\Bigg(\sum_{j=1}^{k} \left\|\partial_{\mu}\vf_{1,j}\right\|_{\infty} + c^*\Bigg)\left\|\vf_{1,k}\right\|_{\infty}^2 + \left\|\partial_{\mu}\vf_{1,k}\right\|_{\infty}\left\|\vf_{1,k}\right\|_{\infty}\Bigg]
\end{equation}
where:
\begin{equation}
c^* \coloneqq {1\over\overline{u}}\left(\chi^{-1}\left(\mu^-\right)+\overline{u}'c\right)
\end{equation}
and $\overline{u}'$ bounds for any $k\in\mathbb{N}^*$ the term $\left\|\left(1-p_{h_0,m_0+\phi_{1,k}}\right)^{-1}\right\|_{\infty}$, so that all the constants which appear in the last expression do not depend on $k$. 
\newline
Let be:
\begin{equation}
\gamma \coloneqq \max\Big\{d,c^*,\overline{u}^2 c^2,3\beta^2\lambda\overline{u}\Big\};
\end{equation}
then, from \eqref{est}:
\begin{equation}
\left\|\partial_{\mu}\vf_{1,k+1}\right\|_{\infty} \le \gamma\Bigg[\Bigg(\sum_{j=1}^{k} \left\|\partial_{\mu}\vf_{1,j}\right\|_{\infty} + \gamma\Bigg)\left\|\vf_{1,k}\right\|_{\infty}^2 + \left\|\partial_{\mu}\vf_{1,k}\right\|_{\infty}\left\|\vf_{1,k}\right\|_{\infty}\Bigg].
\end{equation}
Suppose that for any $1\le j\le k$, $k$ fixed and larger than $1$:
\begin{equation}
\left\|\partial_{\mu}\vf_{1,j}\right\|_{\infty} \le \gamma^{2^j} \ve^{2^{j-1}};
\end{equation}
in this hypothesis:
\begin{equation}
\sum_{j=1}^{k} \left\|\partial_{\mu}\vf_{1,j}\right\|_{\infty} \le \left\|\partial_{\mu}\vf_{1,1}\right\|_{\infty}+ \sum_{j=1}^{k} \gamma^{2^k+1} \ve^{2^k} \le \gamma \ve + \gamma^2\ve\le 2\gamma^2\ve<\gamma
\end{equation}
therefore:
\begin{equation}
\left\|\partial_{\mu}\vf_{1,k+1}\right\|_{\infty} \le \gamma\left(\gamma\gamma^{2^{k-1}}\ve^{2^k}+\gamma^{1\over 2}\gamma^{2^{k-1}}\gamma^{2^k}\ve^{2^{k}}\right)\le \gamma^{2^{k+1}}\ve^{2^k}
\end{equation}
since, by definition:
\begin{equation}
\gamma^{2^{k-1}-2^k+2}+\gamma^{-{1\over 2}} < 1 \qquad \forall k\in\mathbb{N}^*.
\end{equation}
Proposition 5.2 is proved. The result induces the estimate:
\begin{equation}
\left\|\partial_{\mu}m_1-\partial_{\mu}m_0\right\|_{\infty} \le
\sum_{k=1}^{\infty} \left\|\partial_{\mu}\vf_{1,k}\right\|_{\infty} \le 2\gamma^2\ve.
\end{equation}
\qed

\begin{center}
\textit{Global convergence}
\end{center}
We finally prove that also the series of the derivatives converges in the $\alpha$ norm and is bounded by a term of order $\ve$. The strategy is similar to that used in Section 4. 
\newline
Compute:
\begin{eqnarray}
\left|\partial_{\mu}h_{n+1}\left(x\right)-\partial_{\mu}h_{n}\left(x\right) \right| &\le& 
\partial_{\mu}j \ve \int_0^x \left|\chi^{-1}\left(m_{n+1}\left(y\right)\right)
-\chi^{-1}\left(m_{n}\left(y\right)\right)\right|\text{d}y \nn \\
&+& {2\beta j}\ve \int_0^x \left|{m_{n+1}\left(y\right)\partial_{\mu}m_{n+1}\left(y\right)\over\chi^{2}\left(m_{n+1}\left(y\right)\right) }- {m_{n}\left(y\right)\partial_{\mu}m_{n}\left(y\right)\over\chi^{2}\left(m_{n}\left(y\right)\right)}\right|\text{d}y. \nn 
\end{eqnarray}
Since for any $n\in\mathbb{N}^*$ and $x\in[0,\ve^{-1}]$:
\begin{equation}
m_n\left(x\right)\left[1-m_{n\pm 1}^2\left(x\right)\right] < 1,
\end{equation}
there exists $L>0$ which bounds:
\begin{equation}
\left|m_{n+1}\left(x\right)\chi^{2}\left(m_{n}\left(x\right)\right)\partial_{\mu}m_{n+1}\left(x\right)-
m_{n}\left(x\right)\chi^{2}\left(m_{n+1}\left(x\right)\right)\partial_{\mu}m_{n}\left(x\right)\right| \le
L \left|\partial_{\mu}m_{n+1}\left(x\right)-\partial_{\mu}m_{n}\left(x\right)\right|
\end{equation}
so that there is $C'>0$ such that:
\begin{equation}
\left|\partial_{\mu}h_{n+1}\left(x\right)-\partial_{\mu}h_{n}\left(x\right) \right|\le\left(\int_0^x \left|m_{n+1}\left(y\right)-m_{n}\left(y\right)\right|\text{d}y + \int_0^x \left|\partial_{\mu}m_{n+1}\left(y\right)-\partial_{\mu}m_{n}\left(y\right)\right|\text{d}y\right)C'\ve;
\end{equation}
in the $\alpha$-norm:
\begin{equation}\label{ch}
\left\|\partial_{\mu}h_{n+1}-\partial_{\mu}h_{n}\right\|_{\alpha} \le {C'\over\alpha}\Big(\left\|m_{n+1}-m_n \right\|_{\alpha} + \left\|\partial_{\mu}m_{n+1}-\partial_{\mu}m_{n}\right\|_{\alpha}\Big).
\end{equation}
On the other hand, deriving equation \eqref{mn} with respect to $\mu$:
\begin{equation}
\partial_{\mu}m_{n+1}\left(x\right) = \mathscr{U}_{h_{n},m_{n}}
\Big[\chi\left(m_{n+1}\left(x\right)\right)\partial_{\mu}h_{n}\left(x\right)\Big]
\end{equation}
therefore:
\begin{equation}
\left|\partial_{\mu}m_{n+1}\left(x_0\right)-\partial_{\mu}m_{n}\left(x_0\right) \right| \le
\sum_{k=0}^{\infty} \lambda^k \int_{\mathbb{R}^k} \left|\partial_{\mu}h_{n}\left(x_k\right)-\partial_{\mu}h_{n-1}\left(x_k\right) \right| \prod_{j=1}^k J\left(x_{j-1},x_j\right) \text{d}x_j
\end{equation}
which implies by Lemma \ref{lemf}:
\begin{equation}\label{ch1}
\left\|\partial_{\mu}m_{n+1}-\partial_{\mu}m_{n} \right\|_{\alpha} \le {1\over 1-\lambda\mathrm{e}^{\alpha\ve}} \left\|\partial_{\mu}h_{n}-\partial_{\mu}h_{n-1}\right\|_{\alpha}.
\end{equation}
Combining \eqref{ch} and \eqref{ch1} we finally get the recursion:
\begin{equation}
\left\|\partial_{\mu}m_{n+1}-\partial_{\mu}m_{n} \right\|_{\alpha} \le  {C'\overline{u}\over\alpha} \Big(\left\|m_{n}-m_{n-1} \right\|_{\alpha} + \left\|\partial_{\mu}m_{n}-\partial_{\mu}m_{n-1}\right\|_{\alpha}\Big),
\end{equation}
hence, suitably rescaled $\alpha$ with respect its definition \eqref{alphadef}:
\begin{equation}
\left\|\partial_{\mu}m_{n+1}-\partial_{\mu}m_{n} \right\|_{\alpha} \le \left(n+1\right)C'\overline{u} \left({C'\overline{u}\over\alpha}\right)^n \ve
\end{equation}
thus:
\begin{equation}
\sum_{n=0}^{\infty} \left\|\partial_{\mu}m_{n+1}-\partial_{\mu}m_{n}\right\|_{\alpha} \le 
C'\overline{u}\ve \sum_{n=0}^{\infty}  \left(n+1\right)\left({C'\overline{u}\over\alpha}\right)^n
\le O\left(1\right)\cdot\ve.
\end{equation}
This closes the proof of Proposition \ref{p2.5}.
\qed

\subsection{Proof of Theorem 2.1}
It is a matter of collecting the previous results. Call $\ve'$ the biggest value of $\ve$ such that $\det\Delta<1$ and compute:
\newline
\begin{equation}
\left\| m_n-m_0\right\|_{\infty} \le \mathrm{e}^{2\alpha}\sum_{k=0}^{\infty} \left\|m_{n+1} - m_n\right\|_{\alpha} \le C'\overline{u}\mathrm{e}^{2\alpha}\ve\sum_{n=0}^{\infty} \left(C'\overline{u}\over\alpha\right)^n \le O\left(1\right)\cdot\ve.
\end{equation}
Choose then:
\begin{eqnarray}
\alpha &>& \max\left\{{8j\overline{u} \over \delta'\left(2-\delta'\right)},2C'\overline{u}\right\} \\
\ve &<& \min\left\{\widetilde{\ve},\ve',\mathrm{e}^{2\alpha}\right\}
\end{eqnarray}
to guarantee the whole scheme to work.
\newline
Regarding the convergence to the macroscopic solution in the infinite volume limit:
\begin{equation}
\lim_{\ve\,\downarrow\,0}\left\|m-m_0\right\|_{\infty}  \le \lim_{\ve\,\downarrow\,0}\lim_{n\to\infty}
\Big(\|m-m_n\|_{\infty} + \left\|m_n-m_0\right\|_{\infty}\Big) = 0.
\end{equation}
\qed

\section{Conclusions}

The proof of the existence of a smooth, time-invariant profile for the considered Markov stochastic dynamics, represents a first step for a more involved treatment of models in which the interactions are mimicked by long range potentials. The analysis provided here for the peculiar tent shape chosen, moreover, could also apply for a wider class of Kac potentials. 
\newline
We were able to establish a precise connection between the mesoscopic, finite-volume solution and the corresponding profile when the thermodynamic limit is performed, although a proof for the uniqueness of such a solution is missing. Nevertheless, a further analysis should be performed in the case in which a phase transition occurs, i.e. when the boundary magnetizations lie in opposite regions. In this case, the two phases should be connected by an instanton as suggested by numerical results \cite{CDMP}, while a sharp interface appears when passing to the macroscopic scale.

\section*{Acknowledgements}
I am immensely grateful to E. Presutti for his assistance and his valuable hints. I would also like to show my gratitude to A. De Masi and D. Tsagkarogiannis for enlightening discussions and comments that greatly improved the manuscript.

\end{document}